%% file: franzon_proceeding_prerow.tex
\documentclass[12pt]{article}
\usepackage{epsfig}
\usepackage{graphicx}
\usepackage[numbers,sort&compress]{natbib}
\usepackage{subfigure}
\usepackage{amssymb}
\usepackage[english]{babel}
\usepackage{setspace}
\usepackage{graphicx}
\usepackage{amssymb} 
\usepackage{amsfonts}
\usepackage{graphics}
\usepackage{amsmath}
\input econfmacros-csqcd4.tex
\def\Title#1{\begin{center} {\Large {\bf #1} } \end{center}}
\def\be{\begin{equation}}
\def\ee{\end{equation}}
\begin{document}

\Title{Magnetized Neutron Star}
% and Supernovae}

\bigskip\bigskip

%+\addcontentsline{toc}{chapter}{{\it D. Blaschke}}
%+\label{BlaschkeDavid}

\begin{raggedright}

{\it 
Bruno Franzon$^{1}$~~Stefan Schramm$^{1}$%~~and Gerd R\"opke$^{5}$\\
\\
%\thanks{\tt Email: blaschke@ift.uni.wroc.pl}
\bigskip
$^{1}$ Frankfurt Institute for Advanced Studies,
Ruth-Moufang - 1 60438, Frankfurt am Main,
Germany%\\
%\bigskip
%$^{2}$Institut Fizyki Teoretycznej,
%Uniwersytet Wroc{\l}awski,
%pl. Maxa Borna 9,
%50-204 Wroc{\l}aw,
%Poland\\
%
%\bigskip
%$^{3}$Laboratory for Information Technologies,
%Joint Institute for Nuclear Reseaarch,
%Joliot-Curie Str. 6,
%141980 Dubna,
%Russia\\
%
%\bigskip
%$^{4}$Department of Physics,
%Yerevan State University,
%Alek Manukyan Str. 1,
%Yerevan 0025,
%Armenia\\
%{\tt Email: hovik.grigorian@gmail.com}}
%
%\bigskip
%$^{5}$Institute of Physics,
%University of Rostock,
%Universit\"atsplatz 3,
%18013 Rostock,
%Germany\\
%{\tt Email: gerd.roepke@uni-rostock.de}
}

\end{raggedright}

\section{Introduction}

Our main goal in this work is to study magnetized neutron stars  by using a fully general$-$relativity approach presented in the LORENE package\footnote{http://www.lorene.obspm.fr}. Here we have adopted a non-uniform magnetic field profile which depends on the baryon density. This profile has been used in many previous works and seems to be a good choice to explore maximum effects of the internal magnetic field in these objects. Equally important, the magnetic field treated here is poloidal and axisymmetric.   The preliminary results show that stars endowed with a strong magnetic field will be deformed and the mass somewhat increased.

\section{Formalism}
 
The choice of the coordinates in General Relativity is crucial not only to write the gravitational equations in an advantageous form,  but also it can make the problem easier to solve numerically. In the present case,  due to the symmetry of the problem,  a polar$-$spherical type coordinates is chosen, namely,  the Maximal$-$Slicing$-$Quasi$-$Isotropic coordinates (MSQI) \cite{gg}.  The metric in the MSQI coordinates  is written as:
\begin{eqnarray}
ds^2 = -N^{2}dt^{2} + A^{4}B^{2}r^{2}\sin^{2}\theta( d\phi - N^{\phi})^{2}  - \frac{A^4}{B^2} (dr^2 + r^2 d\theta^2)
\label{metric}
\end{eqnarray}
with $N^{\phi} (r, \theta)$ the shift vector, $N (r, \theta)$ the lapse function and A and B are functions of $r$ and $\theta$.  Details of the gravitational equations can be found in the references \cite{bonazzola, bocquet93, bocquet95}. The energy momentum tensor of the system reads:
\be
T_{\alpha\beta} = (e+p)u_{\alpha\beta} + pg_{\alpha\beta} + \frac{1}{4 \pi} \left( F_{\alpha \mu} F^{\mu}_{\beta} - \frac{1}{4} F_{\mu\nu} F^{\mu\nu} \mathrm{g}_{\alpha\beta} \right)
\label{emt}
\ee
where $e$ is the energy density and $p$ the pressure of the fluid. The second term is the electromagnetic contribution and we are  not taking into account the magnetic field in the equation of state. Work along this line is in progress.  Details of the equation of state used in this work can be found in the reference \cite{eos}.

%Carter \cite{carter} shows that the most general electric current for stationary, axisymmetry spacetime  is given by:
%\be 
%j^{\mu} =  \left(j^{t},0,0,j^{\phi} \right).
%\label{current}
%\ee
%It is also shown that the electromagnetic field tensor $F_{\alpha\beta}$ must come from the potential $A_{\alpha}$ which has the property: $A_{\alpha} = A_{t}\mathrm{d}t +  A_{\phi}\mathrm{d}\phi$. 
%The electric field measured by the observer $\mathcal{O}_{0}$ \cite{lich} is:
%\be 
%E_{\alpha} = F_{\alpha\beta}n^{\beta} = 
%\label{electricfield}
%\ee
%$$\left( 0 , \frac{1}{N} \left[  \frac{\partial A_{t}}{\partial r} + N^{\phi} \frac{\partial A_{\phi}}{\partial r}\right ] , \frac{1}{N} \left[  \frac{\partial A_{t}}{\partial \theta} + N^{\phi} \frac{\partial A_{\phi}}{\partial \theta}\right ]   , 0 \right),
%$$
The magnetic field measured by the Eulerian observer is given by \cite{lich}: 
\be 
 B_{\alpha} = -\frac{1}{2} \epsilon_{\alpha\beta\gamma\sigma} F^{\gamma\sigma}n^{\beta} = 
\label{magneticfield}
 \ee
 $$
 \left( 0 , \frac{1}{A^{2} B r^{2} \sin \theta} \frac{\partial A_{\phi}}{\partial \theta}, - \frac{1}{A^{2} B \sin \theta} \frac{\partial A_{\phi}}{\partial r} , 0  \right),
$$ 
where $\epsilon_{\alpha\beta\gamma\sigma} $ is the Levi - Civita tensor  related to the metric $g_{\mu\nu}$ and $n^{\beta}$ the four velocity of the Eulerian observer.  Assuming that the matter inside the star has infinite conductivity, the electric field measured by the comoving observer must be zero.  

%In the references cited above the equation of motion  ($\nabla_{\mu}T^{\mu\nu} = 0	$) can be obtained as:

%\be 
%H \left(r, \theta \right) + \nu \left(r, \theta \right) -ln \Gamma \left( r, \theta \right) + M \left(r, \theta \right) = const,
%\ee
%where 
%\be 
%M \left(r, \theta \right) = M \left( A_{\phi} \left(r, \theta \right) \right): = - \int^{0}_{A_{\phi}\left(r, \theta \right)} f\left(x\right) \mathrm{d}x
%\ee
%\\
%\\
The stress$-$energy tensor of the magnetic field (second term in eq. \ref{emt}) is:
\be 
T^{EM}_{\alpha\beta} = \frac{1}{4 \pi} \left( F_{\alpha \mu} F^{\mu}_{\beta} - \frac{1}{4} F_{\mu\nu} F^{\mu\nu} \mathrm{g}_{\alpha\beta} \right)
\ee
from which one can obtain the sources of the gravitational fields. The energy density reads:
\be 
E^{EM} = \frac{1+ U^2}{8 \pi A^{8} r^{2} \sin^{2}\theta} \left( \partial A_{\phi} \right)^2 ,
\label{energydensity}
\ee
with $U$ the velocity of the fluid in the $\phi$ direction. The momentum density can be written as:
\be 
J^{EM}_{\phi} = \frac{B}{4 \pi} \frac{U}{A^{6} r \sin \theta} \left( \partial A_{\phi} \right)^{2}
\label{momentumdensity}
\ee
and the stress 3-tensor components are given by:

\be 
S^{EM r}_{\;\;\; r} = \frac{1}{8 \pi} \frac{1- U^2}{A^{8} r^{2} \sin^{2} \theta} \left[\left( \frac{\partial A_{\phi}}{\partial r} \right)^{2} - \frac{1}{r^{2}} \left( \frac{ \partial A_{\phi} }{\partial \theta} \right)^2   \right]
\label{stress1}
\ee

\be 
S^{EM \theta}_{\;\;\; \theta} = \frac{1}{8 \pi} \frac{1- U^2}{A^{8} r^{2} \sin^{2} \theta} \left[ \frac{1}{r^2} \left(\frac{\partial A_{\phi}}{\partial \theta} \right)^{2} - \left( \frac{ \partial A_{\phi} }{\partial r} \right)^2   \right]
\label{stress2}
\ee

\be 
S^{EM \phi}_{\;\;\; \phi} = \frac{1}{8 \pi} \frac{1 - U^2}{A^{8} r^{2} \sin^{2} \theta} \left( \partial A_{\phi} \right)^{2}
\label{stress3}
\ee
All the sources term depend on the magnetic vector potential $A_{\phi}$. 
Besides the above definitions,  others important quantities measured by Eulerian obeserver are the $\it{circunferencial \, radius }$ $R_{circ}$
\be 
R_{circ} =  A^{2} (r_{eq}, \frac{\pi}{2})  B ( r_eq,\frac{\pi}{2} ) r_{eq},
\label{rcir}
\ee
being $r_{eq}$ the coordinate equatorial radius, the total gravitational mass of the star
\be 
M =  \int \frac{NA^{6}}{B} (E + S^{i}_{i} + \frac{2}{N} N^{\phi}J_{\phi}) r^{2} \sin\theta dr d\theta d\phi,
\label{mass}
\ee
and an ellipticity which quantifies the apparent oblateness reads

\be
\epsilon = \sqrt{1 -  \left( \frac{r_{p}}{r_{e}} \right)^{2}. }
\label{ellipcity}
\ee

\section{ Magnetic field profile}
Magnetars are know to possess strong magnetic fields which can be estimated from the the period and the period derivative of the star. These highly magnetized neutron stars are believed to have surface magnetic field of order of $10^{14} - 10^{15} $ G .   Besides, the virial theorem states that the magnetic field can be still much higher inside the star than at the surface.  Following the references [8 $-$ 16], a non$-$uniform  magnetic field which can depend on the density is parametrized as:
\be 
B \left(\frac{n_{b}}{n_{0}} \right) =   B_{s} + B_{0} \left[1 - e^{ -\beta \left(  \frac{n_{b}}{n_{0}}  \right)^{\gamma}}\right].
\label{densitymagfield}
\ee
The parameters are the nuclear density at saturation $n_{0}$, the barion denstity of matter $n_{b}$,  the magnetic field on the surface $B_{s}$, and a parameter that controls the magnetic field at the center $B_{0}$.  We set the parameters $\beta$ and $\gamma$ to $\beta = 0.01$ and $\gamma = 2.0$. Other choices for these parameters are possible, but qualitatively the main conclusion remain the same. Usually the value of the magnetic field at the center is about $70 \%$ of $B_{0}$.   Based on the equations defined in the previous sections, in order to construct our models we need to have an expression for the vector magnetic potential $A_{\phi}$.   As we can note in \eqref{densitymagfield} there is no way to get the magnetic potential directly, since the magnetic field is defined as a function of the baryon density $n_{b}$.   Alternatively,   if we suppose a constant magnetic field  in the $z$ direction $\vec{B} = B\vec{z} $, the magnetic vector potential is given by:
\be
\vec{A_{\phi}} = \frac{\vec{r} \times \vec{B}}{2} \approx  \frac{1}{2} r B \sin\theta   \hat{\phi}.
\label{magpotential}
\ee

Our first approximation is to say that the magnetic field $B$ in equation \eqref{magpotential} is locally constant and described by the equation \eqref{densitymagfield}  from which we can easily see that the magnetic field is confined inside the star,  reaches its maximum value at the center of the star.

To illustrate the approach we show on the Figure 1 isocontours of the magnetic field strength,  which linies lie on the surfaces $A_{\phi} = $ const.  
%
%
%
% FIGURE 7
%
%
%
\begin{figure}[htb]
\center
\includegraphics[width=1.2\textwidth,angle=-90,scale=0.5]{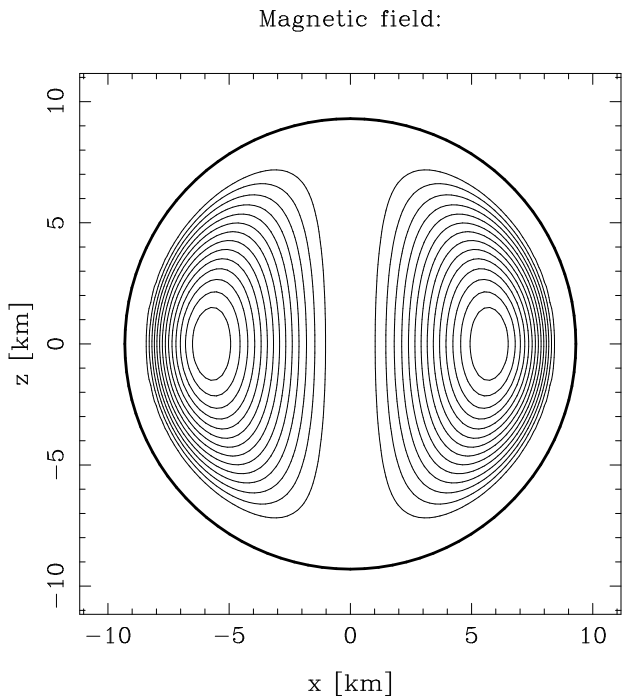}
\caption{\label{Aphi} Isocontours of the magnetic field strength in the (x, y) plane}
\end{figure}
%%%%%%%%%%%%%%%%%%%%%%%%%%%%%%%%%%%%%%%%%%%%%%%%%%%%%%%%%%%%%%%%%%%%%%%%%
%%
%%   use this format to include an .eps figure into your paper
%%
%\begin{figure}[htb]
%\includegraphics[width=0.6\textwidth]{Delta_Pauli}
%\includegraphics[width=0.6\textwidth]{Mott-Density}
%\caption{Binding energy $E_{B}$ and continuum edge $u(p=0,P_{F})$ for nucleons
%in symmetric nuclear matter (solid lines) and pure neutron matter (dashed
%lines) as a function of the density at $T=0$. 
%The nucleon dissociation (Mott transition) occurs at the densities where the 
%binding energy merges the corresponding continuum edge.
%Left panel: The role of the Pauli blocking for the Mott effect is shown.
%Right panel: The Mott densities are almost independent of the choice of the
%potential parameter $V_0$, see also Table \ref{tab:parameters}.}
%\label{fig:Mott}
%\end{figure}
%%%%%%%%%%%%%%%%%%%%%%%%%%%%%%%%%%%%%%%%%%%%%%%%%%%%%%%%%%%%%%%%%%%%%%%%%%%
For a specific choice of central energy density we have obtained a mass  of $M = 1.31 \,M_{sun}$ and a circunference radius of $R_{circ} = 11.34$ km, whose result are  very close to the spherical case and without magnetic field.  The ellipticity was found to $\epsilon = 0.032$.
%The same model, but non magnetized,  we have found  $M = 1.31  \,M_{sun} $  for the mass and $R_{circ} = 11.35$ Km for the circumference radius. The ellipticity ellipticity was found to $\epsilon = 0.032$. 
This tiny effect is due to $\it{low}$ value of the magnetic field, namely, $B = 3.5\times 10^{17}$ G  at the center,  which is not enough to deform the star. In order to investigate how the mass and the ellipticity, and therefore, the deformation, change with the magnetic field we have evaluated models for different values of $B_{0}$.
 
Figure 2 shows the change in the mass, while the Figure 3 presents the ellipticity as a function of the magnetic field. 
%\begin{figure}[h]
%\center
%\subfigure[ref1][Curve of the gravitational mass $M$ as a function of central magnetic field.]{\includegraphics[width=0.6\textwidth,angle=-90,scale=0.5]{mass_bfield.eps}}
%\qquad
%\subfigure[ref2][ Curve of the ellipticity $\epsilon$ as a function of the central magnetic field.]{\includegraphics[width=0.6\textwidth,angle=-90,scale=0.5]{ellipticity_bfield.eps}}
%\caption{Global properties of the neutron star as a function of the magnetic field.}

%\end{figure}
%
% FIGURE 8
\begin{figure}[h]
\center
\includegraphics[width=1.0\textwidth,angle=-90,scale=0.5]{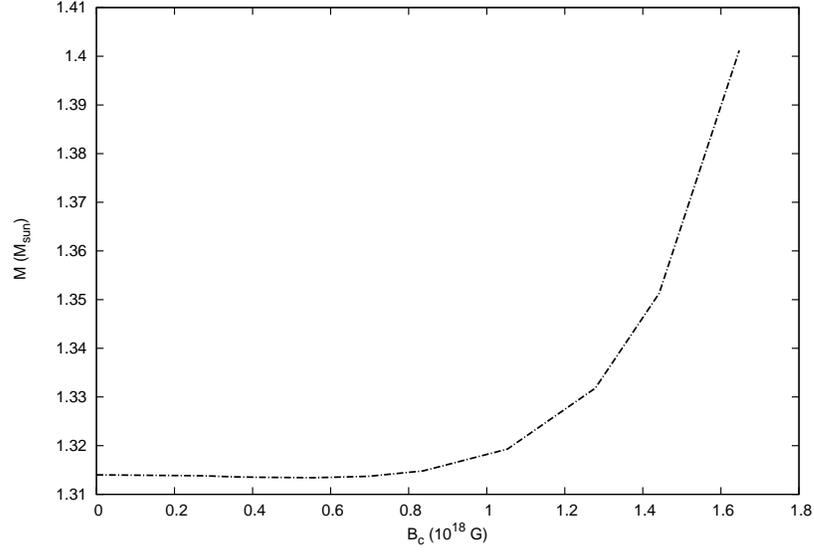}
\caption{\label{fieldmass} Curve of the gravitational mass $M$ as a function of central magnetic field. } 
\end{figure}

\begin{figure}[!h]
\center
\includegraphics[width=1.0\textwidth,angle=-90,scale=0.5]{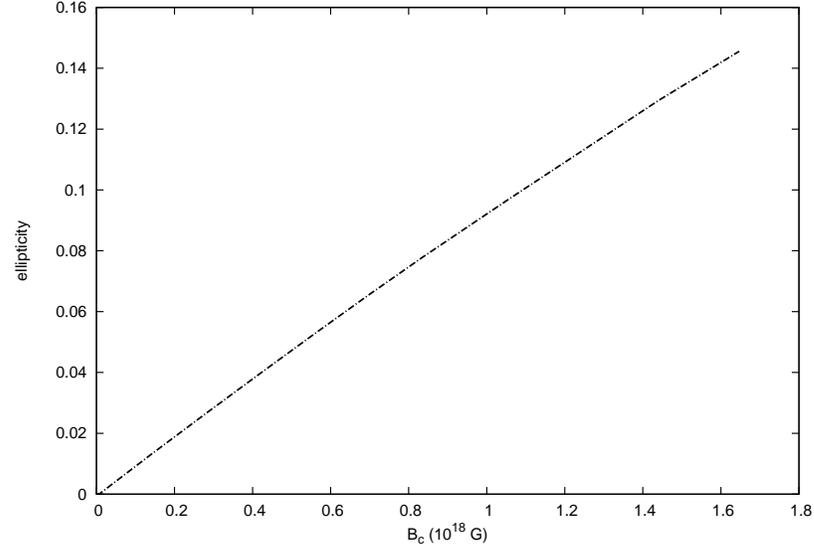}
\caption{\label{fieldmass}  Curve of the ellipticity $\epsilon$ as a function of the central magnetic field.} 
\end{figure}

%
%
%
% FIGURE 9
%
%
%\begin{figure}[!h]
%\center
%\includegraphics[width=0.6\textwidth,angle=-90,scale=0.5]{ellipticity_bfield.eps}
%\caption{\label{fieldellipticity} Curve of the ellipticity $\epsilon$ as a function of the central magnetic field. }
%\end{figure}
%
%
From these figures,  both the mass and the ellipticity increase with the magnetic field whose maximum value corresponds to a value of $1.64\times 10^{18}$ G at the center of the star. The corresponding ellipticity is found to be $\epsilon = 0.146$, showing that the deformation plays an important role in the global properties of the star. According to the Fig.2 the increasing in the mass is about $7\%$,  less than previous calculation by solving TOV equation, and higher than the $2\%$ predicted by the perturbative approach  \cite{ritam}.

\section{Conclusion}
 We have used the $Lorene\,\, package$ for computing perfect fluid magnetized stars in general relativity with the inclusion of a non-uniform magnetic field profile.  We have then used the code to see the effect of the magnetic field strength on the deformation and also on the mass of the star.  The maximum magnetic field found at center is around $1.64\times10^ {18}$G for a fixed stellar mass of $M = 1.40\,M_{sun}$.  The results were obtained for the APR equation of state. A study with a more sophisticated equation of state including magnetic effects  and more realistic field configurations is in progress.
%\begin{acknowledgments}
%\end{acknowledgments}

%%%%%%%%%%%%%%%%%%%%%%%%%%%%%%%%%%%%%%%%%%%%%%%%%%%%%%%%%%%%%%%%%%%%%%%%%%%

\subsection*{Acknowledgement}
The authors thanks to Veronica Dexheimer, Joachim Frieben and D. Chatterjee  for fruitful comments and discussions. Bruno Franzon wishes to thank the financial support of the CNPq and DAAD.

\end{document}

%% file: econfmacros-csqcd4.tex
\def\beq{\begin{equation}}
\def\eeq#1{\label{#1}\end{equation}}
\def\eeqn{\end{equation}}

\def\beqa{\begin{eqnarray}}
\def\eeqa#1{\label{#1}\end{eqnarray}}
\def\eeqan{\end{eqnarray}}

\let\bar=\overbar

\def\Dslash{\not{\hbox{\kern-4pt $D$}}}
\def\dslash{\not{\hbox{\kern-2pt $\del$}}}

\def\ee{e^+e^-}

\def\msb{{\bar{\ssstyle M \kern -1pt S}}}

\usepackage{fancyhdr,graphicx}